\documentclass[lettersize,journal]{IEEEtran}

\usepackage{cite}

\usepackage{amsmath}
\usepackage{algorithm}
\usepackage{algorithmic}
\usepackage{subfigure}
\usepackage{amssymb}
\usepackage{graphicx}
\usepackage{verbatim}
\usepackage{empheq}

\usepackage{array}
\usepackage{mathtools}
\usepackage{graphicx}
\usepackage{cases}
\usepackage[colorinlistoftodos]{todonotes}
\usepackage[colorlinks=true, linkcolor=black,citecolor=black,urlcolor=black]{hyperref}
\usepackage{indentfirst}
\usepackage{multirow}
\usepackage{booktabs}
\usepackage{makecell}

\allowdisplaybreaks

\begin{document}
\title{Anchor-and-Connect: Robotic Aerial Base Stations Transforming 6G Infrastructure}
\author{
Wen~Shang, Yuan~Liao, Vasilis~Friderikos,  Halim~Yanikomeroglu

\thanks{Wen Shang, Yuan Liao \&  Vasilis Friderikos, are with the Department of Engineering, King's College London, London WC2R 2LS, U.K. (e-mail: wen.shang@kcl.ac.uk; yuan.liao@kcl.ac.uk; vasilis.friderikos@kcl.ac.uk).

Halim Yanikomeroglu is with Non-Terrestrial 
Networks (NTN) Lab, Department of Systems and Computer Engineering, Carleton University, Ottawa, ON K1S 5B6, Canada (e-mail: halim@sce.carleton.ca).}
}


\maketitle

\begin{abstract}

Despite the significant attention that aerial base stations (ABSs) have received recently, their practical implementation is severely weakened by their limited endurance due to the battery constraints of drones. To overcome this fundamental limitation and barrier for wider adoption, we propose the concept of robotic aerial base stations (RABSs) that are equipped with energy-neutral anchoring end-effectors able to autonomously grasp or perch on tall urban landforms. Thanks to the energy-efficient anchoring operation, RABSs could offer seamless wireless connectivity for multiple hours compared to minutes of the typical hovering-based ABSs. Therefore, the prolonged service capabilities of RABSs allowing them to integrate into the radio access network and augment the network capacity where and when needed. To set the scene, we discuss the key components of the proposed RABS concept including hardware, workflow, communication considerations, and regulation issues. Then, the advantages of RABSs are highlighted which is followed by case studies that compare RABSs with terrestrial micro BSs and other types of non-terrestrial communication infrastructure, such as hovering-based, tethered, and laser-powered ABSs.
\end{abstract}



\section{Introduction}
\label{introduction}

Aerial base stations (ABSs) mounted on airborne platforms, such as uncrewed aerial vehicles (UAVs), exhibit an increasing potential within the developing landscape of 6G cellular networks thanks to their functional flexible deployment and three dimensional maneuverability. Summarized by standards organization including the 3rd Generation Partnership Project (3GPP) and the International Telecommunication Union (ITU), ABSs could be utilized to play different roles in wireless systems, such as small base stations (BSs), mobile relays, data collectors, and moving anchors for localization. Although ABSs have attracted significant attention from both academia and industry in recent years, the limited endurance restricted by the capacity of onboard batteries is still a critical issue in sustainable networks. This issue becomes particularly pronounced when ABSs are intended to offer daily mobile data service rather than just being used for very short-term sporadic emergency communications. With this in mind, we are exploring the use of robotic aerial base stations (RABSs) with anchoring capabilities via robotic manipulators \cite{friderikos2021airborne} to allow for a massive improvement in providing long-term wireless connectivity compared to hovering/flying ABSs. More specifically, by taking advantage of the grasping capabilities \cite{liu2019vision}, RABSs can attach autonomously to lampposts or other urban landforms to serve as small cells for multiple hours or even longer. This is in contrast to the flying/hovering endurance for the small-size rotary-wing UAVs acting as small BSs which is significantly less than an hour. 


\begin{figure*}[!t]
\centering
\includegraphics[width=0.95\linewidth]{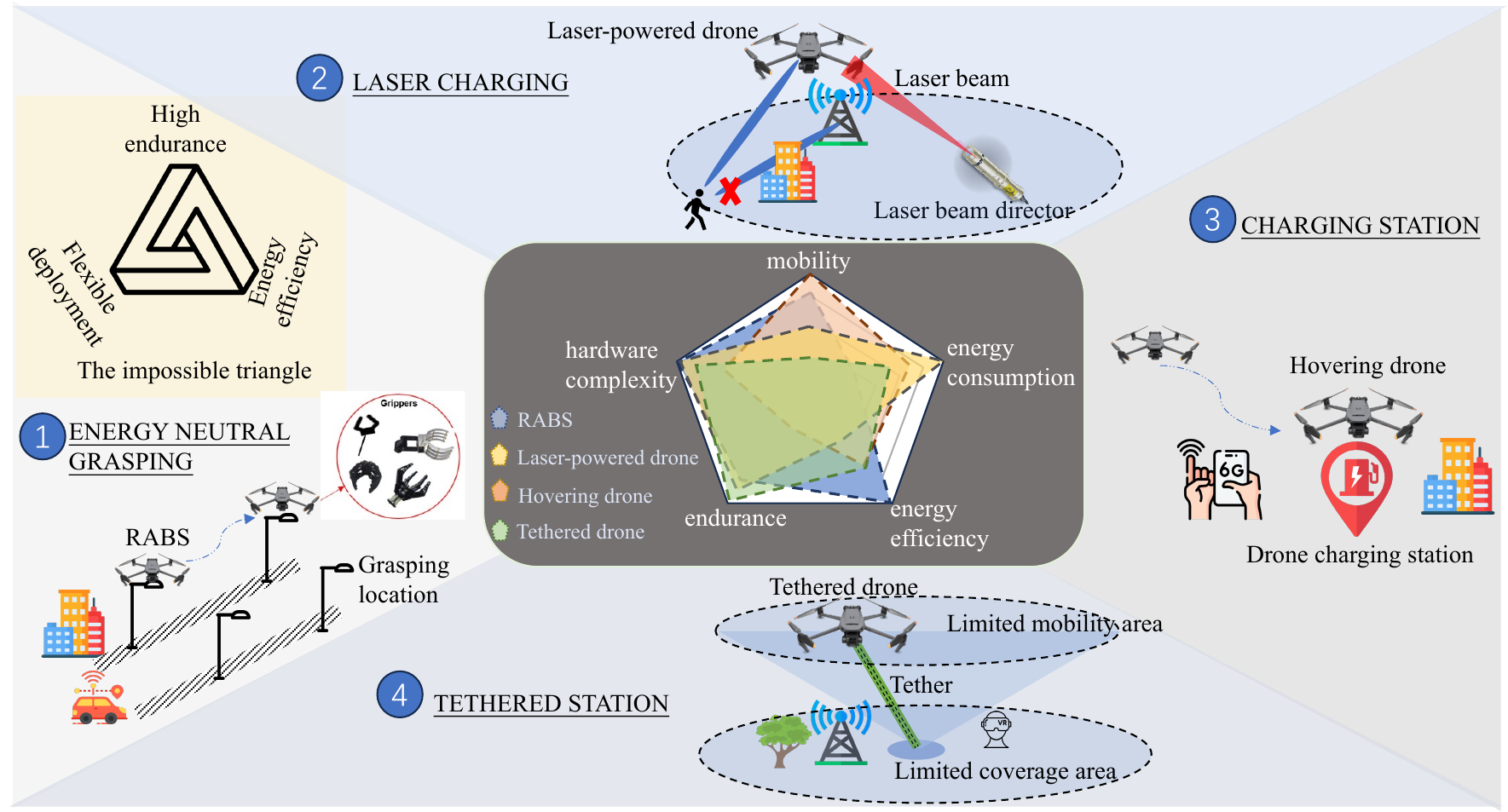}
\caption{An overview of different ABS platforms developed to overcome the endurance issue: RABS, laser-powered ABS, hovering ABS, and tethered ABS.}
\label{scenario}
\end{figure*}

Recently, significant efforts have been devoted to addressing the endurance issue of aerial base stations since this is a major limiting factor. Figure \ref{scenario} depicts an overview of different ABS platforms developed to overcome the endurance issue. The wireless power transfer technique represents a promising solution for powering ABSs without service interruption, allowing for recharging at ground stations. In \cite{lahmeri2022charging,lahmeri2022laser}, high-energy laser beams transmitted from a ground station are utilized to power ABSs. Theoretically, the laser-powered ABS can provide an unlimited serving time when in close proximity to the charging station; however, it cannot be considered as a scalable solution in terms of energy efficiency and it is also highly problematic in terms of safety for dense urban environments. In \cite{alzenad2018fso}, the free space optics (FSO) beam is proposed to recharge ABSs via  wireless power transfer. However, the energy transfer efficiency of FSO is greatly affected by environmental factors, e.g., when an ABS is located at an altitude of 1 km, the link margin is 38 dB in clear weather in contrast to -15 dB in a foggy day. The authors of \cite{zhang2023deployment} utilize a tethered UAV to carry an ABS, in which a tether linking the UAV with a ground station could transfer both data and power simultaneously. Although tethered UAVs can communicate with ground devices without the need to recharge, this seamless connection comes at the expense of limited flexibility. 
Moreover, Nokia Bell Labs developed a landing-based ABS prototype named F-cell \cite{bajracharya20226g}, to satisfy emergency communication requirements. The F-cell can be carried and transported by a UAV, and then dropped on rooftops or other easy-to-land platforms to offer wireless service. Each F-cell is equipped with a solar panel to provide the processing energy and a 64-antenna array to enhance the wireless backhaul capacity. However, the F-cell requires a flat area to take off and land, thereby significantly restricting its operational capabilities, whilst facing safety challenges since it is not fixed to the surface. Notably, RABSs can be deemed as autonomous robots that are able to safely grasp in wide set of different tall urban landforms by utilizing robotic manipulators. Besides, unlike the F-cell that is transported and relocated by a UAV, RABS can respond to communication requirements more flexibly and quickly thanks to its inherent flying ability. 

\begin{figure}[!t]
\centering
\includegraphics[width=0.9\linewidth]{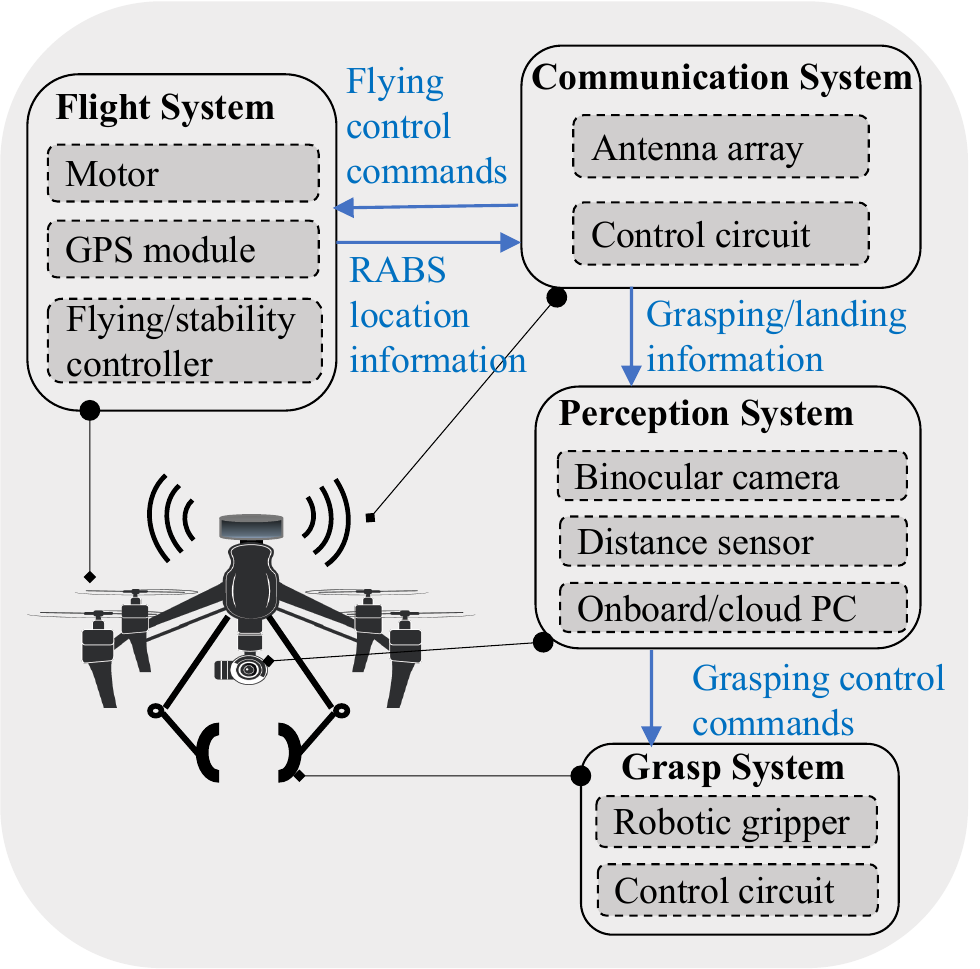}
\caption{Hardware structure of the designed RABS.}
\label{RABS_architecture}
\end{figure}
\textcolor{black}{Conventional drone-based stations typically remain airborne for only 30 to 45 minutes, which limits their use primarily to niche scenarios such as emergency and rescue operations. In contrast, extending operational times to several hours or even days would unlock a wide range of new opportunities.} Considering the functional movable ability and energy efficiency, we envision a wide range of possible use cases for RABSs in dense urban areas to assist towards the aim of network densification \textcolor{black}{without the need for permanent infrastructure deployment. }
Firstly, considering the low data rate of cell-edge users and the high possibility of BSs being overloaded during unexpected crowded events, i.e., congestion episodes, RABSs can be deployed to track spatial-temporal traffic variations as small BSs or relays, to significantly enhance the network connectivity and extend the cell coverage effectively. Secondly, in next-generation cellular networks, new radio applications always place more extensive and stringent requirements on communication systems, in which cases RABSs can be utilized to satisfy these new radio applications, e.g., RABSs attached to lampposts can act as roadside units to offer low-latency and high-reliable service in intelligent transportation systems. They can also be installed with edge computing or caching modules to alleviate the overload and congestion in cloud servers in case of congestion episodes. Thirdly, RABSs can cooperate with other types of non-terrestrial nodes located at different vertical platforms, such as high-altitude platform stations (HAPs) as well as satellites, and provide wireless access from the side closer to ground users. Lastly, with full control and longer endurance, RABS with perception systems can employ integrated sensing and communication technology to sense and manage specific information, such as capturing crop information and assisting decisions in smart agriculture. 

\section{System Architecture Design Considerations and Challenges}
\label{SystemArchitecture}

\subsection{A Primer on Hardware Design}
\label{RABSPrototype}

The recent development in multi-rotor UAVs enables ABSs to perform a wide range of operations in a controllable manner, i.e., flying along a predefined route, hovering at a specific location and cooperating with other ABSs. According to their mechanical design, UAVs can be classified into fixed-wing and rotary-wing UAVs. Because small-size rotary-wing UAVs are able to take off and land on various types of urban platforms and yield more flexible operations, they can be chosen to carry and transport the RABSs. The hardware design of RABS is illustrated in Figure \ref{RABS_architecture}, which includes four subsystems to perform different operations, i.e., flight, communication, grasp, and perception systems. We propose the designed structure and operations of these subsystems. 

Unlike conventional ABSs, a robotic gripper is mounted on the RABS so that it can attach to different urban landforms, such as lampposts and rooftops in a flexible and safe manner. In \cite{zhang2020state}, robotic grippers are classified into several categories via different classification strategies. Specifically, grippers can be divided into four categories according to the number of fingers, i.e., 2-finger gripper, 3-finger gripper, 4-finger gripper, and anthropomorphic hand. The finger design is always determined by the contacting surfaces with the objects when performing the grasping operation. Inspired by the fact that nearly 60\%-70\% of human grasping actions of objects with parallelepiped, pyramidal, and cylindrical shapes can be operated by two fingers, the 2-finger gripper is the most popular design for practical industrial applications \cite{zhang2020state}. According to the type of actuation used, grippers can be classified as vacuum, magnetic, hydraulic, pneumatic, and electric types. Considering the RABS is battery-powered and the contacting surface of typical landforms (e.g., lampposts) is usually metal, magnetic or electromagnetic grippers might be employed by RABSs as well. The required force to hang the RABS to a lamppost via a 2-finger gripper is calculated by the Eq. (1) in \cite{friderikos2021airborne}. Taking the DJI Matrice 300 RTK UAV with a weight of 6.3 kg as an example, the force requirement from the gripper should be 1234.8 N in the worst case, \textcolor{black}{and the maximum flying time is 55 minutes (without payload) or 31 minutes when it is at its maximum weight of 9 kg.
The integration of grippers enhances anchoring stability and reliability but increases the overall payload. Since rotary-wing UAV propulsion power scales approximately linearly with mass, every added weight consumes additional power and reduces endurance. Therefore,} this added weight of the gripper and other subsystems needs to be carefully designed, to balance the overall load capacity and power consumption of the RABS which can limit the service performance.

\subsection{Workflow of Robotic Aerial Base Stations}

To deploy RABSs at suitable locations to adapt to the traffic dynamic in both spatial and temporal domains, a central controller needs to monitor the changes of traffic in real-time, and timely send commands to RABSs that they need to relocate their positions. Once the RABS receives the control signal from the ground control stations and decides to grasp at a certain location, it is required to fly to this target point and grasp in a reliable and precise way. To achieve this task, RABS should be equipped with a variety of onboard sensors to control its localization, velocity, and attitude via a global navigation satellite system and radar-based localization techniques. Furthermore, to complete the grasping function at a chosen location autonomously, real-time object detection capability is required in the RABS. Fortunately, thanks to the recent development of deep learning methods, UAVs can perform autonomous and real-time grasping precisely via vision-based and AI-aided approaches. We take the UAV with grasping ability designed in \cite{liu2019vision} as an example. For the hardware design, a computer deployed in the ground station is used to offer enough computational capacity for vision measurements, and a downward-looking binocular camera is mounted onboard to recognize the grasping target. For the software structure, the Yolo3 algorithm is employed to perform the object detection, which includes three steps, i.e., collect and label samples, train the neural network, and test the trained model. The developed UAV platform is then able to perform the grasping task accurately. Note that since RABSs will be part of the 6G and beyond network infrastructure, those computationally heavy tasks could potentially be offloaded to suitable edge clouds instead of an extra computer. \textcolor{black}{To compromise the trade-off between the onboard processing and reliance on edge-cloud resource, latency critical tasks such as vision-based detection can be allocated on-board, while computationally heavily but delay tolerant tasks, such as neural network training, can be offloaded to the edge.} The allocation of capacities must balance the competing needs of edge computing task offloading and traffic demand satisfaction in hotspot areas.

\textcolor{black}{Despite the availability of advanced object detection techniques and AI-based real-time inference, identifying stable grasping points in urban infrastructure remains non-trivial. Visual features can be degraded by illumination changes, shadows, adverse weather such as fog or rain. Furthermore, safe and efficient trajectories must account for buildings, cables, and other urban obstacles in urban airspace. Motion planning for RABS must therefore not only guarantee obstacle avoidance but also ensure alignment of the gripper with candidate anchoring locations. The last stage of grasping is particularly sensitive to environmental disturbance such as wind gusts and turbulence. Consequently, robust control approaches capable of adapting to dynamic conditions are essential to achieve secure and reliable . 
}
\subsection{Wireless Access and Backhaul Link}
\label{WirelessLink}

In this subsection, we discuss the spectrum design for both access and backhaul links of RABSs. For the access link, RABSs can communicate on high-frequency bands, such as mmWave or sub-THz, to support extremely high-capacity networks via network densification and ultimately enable novel applications with multi-modalities. Operating RABSs on high-frequency bands has the following two benefits. Firstly, it could avoid interference with the existing communication systems operating on low frequencies. Secondly, because carriers on these high frequencies are always characterized by very high path loss in non-line-of-sight (NLoS) conditions, RABSs can be deployed carefully to improve line-of-sight (LoS) probability due to their high flexibility. 
On the other hand, unlike terrestrial BSs connecting with the core network through a high-capacity fiber link, RABSs require wireless backhaul due to their frequent movement. We are proposing two potential approaches to enhance the backhaul capacity. Firstly, RABSs can connect with the macro BSs on high-frequency bands to guarantee the backhaul requirements. Integrated access and backhaul (IAB), a concept proposed by 3GPP standards to flexibly support wireless access and backhaul links with the same infrastructure, can be employed for RABSs-assisted networks with reduced hardware costs \cite{shang2024unlocking}. Numerical results in the study \cite{shang2024unlocking} demonstrated that RABSs-assisted networks with optimized spectrum management and anchoring locations can support access and backhaul links effectively with promising performance. 
Secondly, by mounting a multi-antenna array on RABSs, backhaul link capacity can be improved via MIMO techniques, which have been applied in the Nokia F-cell. 
\begin{figure*}[!t]
\centering
\includegraphics[width=0.95\linewidth]{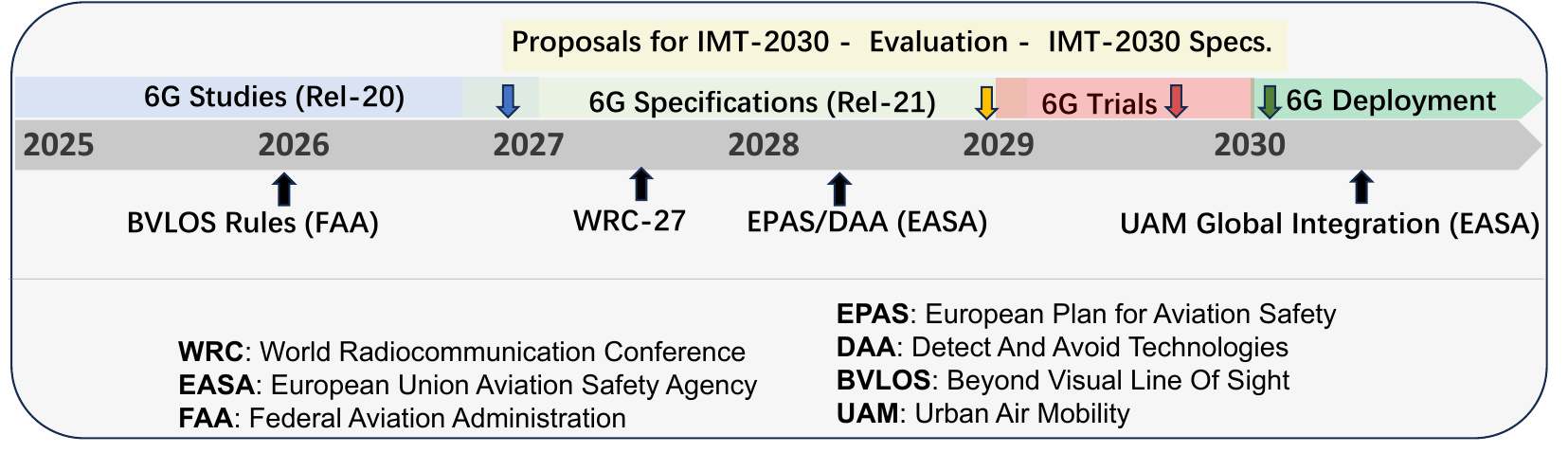}
\caption{Standardization efforts planned for 6G wireless communication networks and efforts to regulate UAV operation.}
\label{standrds}
\end{figure*}

\subsection{Standardizations and Regulatory Aspects}
Implementing ABS-assisted communication networks requires adherence to both standardizations and regulatory aspects.
Organizations such as the 3GPP, IEEE, and the International Telecommunication Union Telecommunication Standardization Sector (ITU-T) identify UAV-specific standards, including performance requirements, technologies, and protocols \cite{bajracharya20226g}. 
For example, to enable efficient network performance, the application of spectrum and corresponding channel modeling for aerial platform operations need to be supported by related standards. 
At the same time, in a rather independent way there are international efforts to regulate UAV operations that involve multiple organizations, treaties, and national policies aimed at ensuring safe and responsible UAV use. 
Fig. \ref{standrds} presents an overview timeline of standardization efforts planned by IMT-2030 as well as UAV specific regulations.
It is particularly noteworthy that by 2030, coinciding with the anticipated initial rollout of 6G networks, there will be a global integration of Urban Air Mobility (UAM) aimed at facilitating airspace management, regulatory harmonization, infrastructure development, and public acceptance.
RABS operation could present a new interference use case, necessitating the need for explicit standardization to enable efficient integration.
\textcolor{black}{For the efficient incorporation of the proposed RABSs into networks, the following provide potential aspects for standardization to be specified into 3GPP Rel-20 and beyond:
\begin{itemize}
  \item Perching State and Flight Path Reporting: Enhance flight path reporting to include real-time updates on perching locations and durations for grasping. Introduce new Information Elements (IEs) for “perching state” (e.g., attached, detached, transitioning). This would enable serving gNBs to anticipate dynamic changes in coverage, interference patterns, and resource allocation during RABS relocation and anchoring.
  \item RABS as a Distinct Category of UxNB: Updates to include RABS as a distinct category of UxNB (UAV-based gNB). Standardized requirements would include procedures for dynamic cell activation and deactivation based on perching status, as well as signaling support for handovers triggered by relocation. 
  \item Integration with UTM and Infrastructure Registry: Extend requirements for real-time perching location reporting to Unmanned Aerial Traffic Management (UTM), including lamppost-specific identifiers (e.g., GPS coordinates or municipal asset IDs), standardized in a secure and verifiable format. This would allow regulators and network operators to maintain a trusted registry of authorized perching points, reducing the risk of unauthorized deployments and ensuring safe coexistence with urban infrastructure.
\end{itemize}}

The use of aerial platforms such as UAVs is governed by strict regulations aimed at ensuring operational safety. A primary motivation for these regulations is to minimize risks to piloted aviation and mitigate risks to life or property from falling debris. For example, in the UK, UAVs must fly below 120 m above local ground level and away from Flight Restrictions Zones, such as airports and spaceports. These restrictions should be straightforward to meet without significantly affecting the utility of RABS. More problematic is the requirement to always fly more than 50 m horizontal distance from people, including those inside buildings, in the case of UAVs above 250 gr in weight. It is acceptable for lighter UAVs to fly above people, but this will be hard to attain for practical RABS. Clearly, this area of regulation will need to evolve for RABS to achieve potential in offering capacity enhancements in populated areas. Within that context, the successful deployment of RABSs heavily depends on the establishment of low-risk aerial urban corridors to enable ubiquitous roaming of aerial small cells within an urban environment. Since the same regulatory challenges are faced by industries such as drone delivery, it is expected that through responsible development, those aspects will be utilized. 

\section{Ideal Attributes of Robotic Small Cells}\label{Advantages}

In this section, we discuss key attributes of RABS-assisted wireless systems that can be deemed as advantages compared with other aerial platforms used for communications.

\textbf{\textit{1) Overcoming the ABS Endurance Issue:}} As one of the initial motivations, the serving endurance of RABSs can be multiple hours or even longer, compared to the hovering/flying time of conventional ABSs which is always less than one hour. 
Derived from the power model shown in work \cite{zeng2019accessing}, the flying power value that can maximize flying endurance is approximately 100 W for a fixed-wing UAV and 160 W for a rotary-wing UAV, while the hovering power is around 170 W. Note that flying power can be up to several hundred watts, depending on the UAV's characteristics. \textcolor{black}{Environmental factors such as wind gusts can further increase energy consumption.}
Comparing the hovering, flying, and communication power, it can be observed that the propulsion power is much more significant than communication-related counterpart during the ABS serving endurance. Moreover, noticing that the grasping power, normally several watts, is significantly less than the hovering and flying power, RABSs, which provide service when grasping at urban landforms, certainly have a longer lifetime than conventional ABSs. Most importantly, it should be noted that the specialized dexterous grippers can be designed to operate in an energy-neutral manner, in which case the perching energy consumption can be completely eliminated. \textcolor{black}{In addition, communication system efficiency, which becomes the main power consumption during anchoring, can be improved to further extend service endurance. Advanced techniques, such as hybrid beamforming, can be employed to reduce RF power consumption while maintaining high data rates.}

\textbf{\textit{2) Enhanced Object Detection and Tracking Including Sensing:}} 
Owing to their high mobility, ABSs are employed in various important applications in addition to communication. UAVs equipped with cameras can deliver more efficient and convenient performance in computer vision than fixed-location surveillance cameras \cite{du2018unmanned}, enabling capabilities such as integrated sensing and communication (ISAC), 
However, this mobility introduces new challenges. A primary challenge arises from the high altitude of the traditional UAV, typically above 50 meters. At such altitudes, vehicles and pedestrians appear small and occupy very few pixels per frame, making object recognition difficult. In contrast, the proposed RABSs, which can anchor to urban landforms such as lampposts and rooftops, typically between 5 m to 15 m, offering an advantage by staying stationary at low altitudes. This capability enables them to perform advanced vision tasks.
The study in \cite{du2018unmanned} investigated fundamental computer vision tasks, including object detection and tracking, using the dataset captured by UAVs operating at varying altitudes across complex scenarios. Specifically, three altitude levels were examined: low-altitude (10-30 m), medium-altitude (30-70 m), and high-altitude ($>$ 70 m). The numerical evaluations, conducted with a range of detection and tracking algorithms, demonstrated a significant impact of the flying altitude on performance. For instance, compared to high-altitude views, low-altitude views improved object detection precision by approximately 40\% and tracking accuracy by approximately 20\%. This improvement is attributed to the enhanced detail of objects captured at lower altitudes and the reduced clutter in the background. Therefore, the proposed RABS, anchoring at low-altitude urban structures, can perform enhanced sensing tasks with higher clarity view and longer endurance.

\textbf{\textit{3) Network Densification Without Densifying the Infrastructure:}} To deal with the rapid growing volume of traffic demand, one promising technique is network densification, i.e., deploying a large number of small cells densely. However, because the traffic distribution in urban regions shows high inhomogeneity in both spatial and temporal domains, network densification via densifying the wireless nodes may cause significant waste in both capital and operating expenditure. In contrast, RABS provides an alternative approach for low-cost and flexible network densification. More specifically, RABSs could be deployed in hot spot regions and transported to other peak traffic areas by following the temporal traffic dynamic as it unfolds. In that sense, network densification via RABS could ensure capacity is always targeted at the specific locations and times when it is most needed. In other words, compared with terrestrial small BSs, RABS show a higher flexibility that allows them to follow the spatio-temporal traffic variations.


\textbf{\textit{4) Adapting to Harsh Weather Conditions and Low-noise Operation:}} One of the most critical issues when operating UAVs is harsh weather conditions, such as rainstorm, wind, and extreme temperature environments. Specifically, heavy rain, snow, and other water might stop the onboard motors and threaten the UAV flight safety. For example in 2020, the UK had 170 days in which 1 mm or more of rainfall, covering $50\%$ of the days. 
Flying or hovering in a windy environment consumes more energy to control the UAV and sometimes is even infeasible to take off. Furthermore, vision-based operations, such as monitoring and sensing, might be affected in foggy weather. In most of these cases, the grasping-based RABSs could provide service in a much safer manner due to the anchoring capability, which can be summarized as rainproof and windproof ability.  

The number of UAV-assisted applications has grown significantly in recent years, such as cargo delivery, environmental monitoring, smart agriculture, and wireless communication. Within that context, UAVs are regarded as a new source of environmental noise pollution and are starting to attract increasing attention from researchers. As shown in table \ref{Noise}, the authors of \cite{alexander2019flyover} measure the noise emission of DJI Matrice 600 Pro when flying and hovering. It can be seen that the noise generated by the UAV operating at a low altitude is very close to the so-called acceptable-noise-level (ANL) which is 85 dB. Furthermore, recalling that the free-field sound decay rate is at approximately 6 dB with a doubling of the distance, the UAV operating at the highest permitted altitude of 120 m would still produce nearly 50 dB of noise pollution. Note that the limits for urban ambient noise are at 45 dB during the night and at 55 dB during the day. Therefore, noise pollution is a critical and real issue when applying ABSs in urban areas, especially when considering public acceptance issues. However, when RABSs grasp at lampposts and switch-off their rotors, they can provide zero levels of noise emission and are thus more suited for urban applications. 

\begin{table}[!t]
\centering
\caption{Noise Emission of DJI Matrice 600 Pro \cite{alexander2019flyover}}
\label{Noise}
\begin{tabular}{p{2 cm} | p{1.4 cm} p{2.3 cm} | p{1.4 cm}}
\hline
UAV status & Altitude (m) & Lateral distance (m) & Noise (dB)\\
\hline
\multirow{5}*{Hovering} & \multirow{5}*{9.18} & 0 & 89.6 \\
~ & ~ & 2.45 & 88.7 \\
~ & ~ & 5.28 & 87.3 \\
~ & ~ & 9.14 & 83.7 \\
~ & ~ & 15.84 & 79.2 \\
\hline
\multirow{5}*{Flying (3.23 m/s)} & \multirow{5}*{7.5} & 0 & 85.3 \\
~ & ~ & 2.45 & 84.0 \\
~ & ~ & 5.28 & 82.7 \\
~ & ~ & 9.14 & 79.6 \\
~ & ~ & 15.84 & 75.9 \\
\hline
\end{tabular}
\end{table}

\section{Case Study and Investigation}
\label{casestudy}

\begin{figure*}[!t]
\centering
\includegraphics[width=1\linewidth]{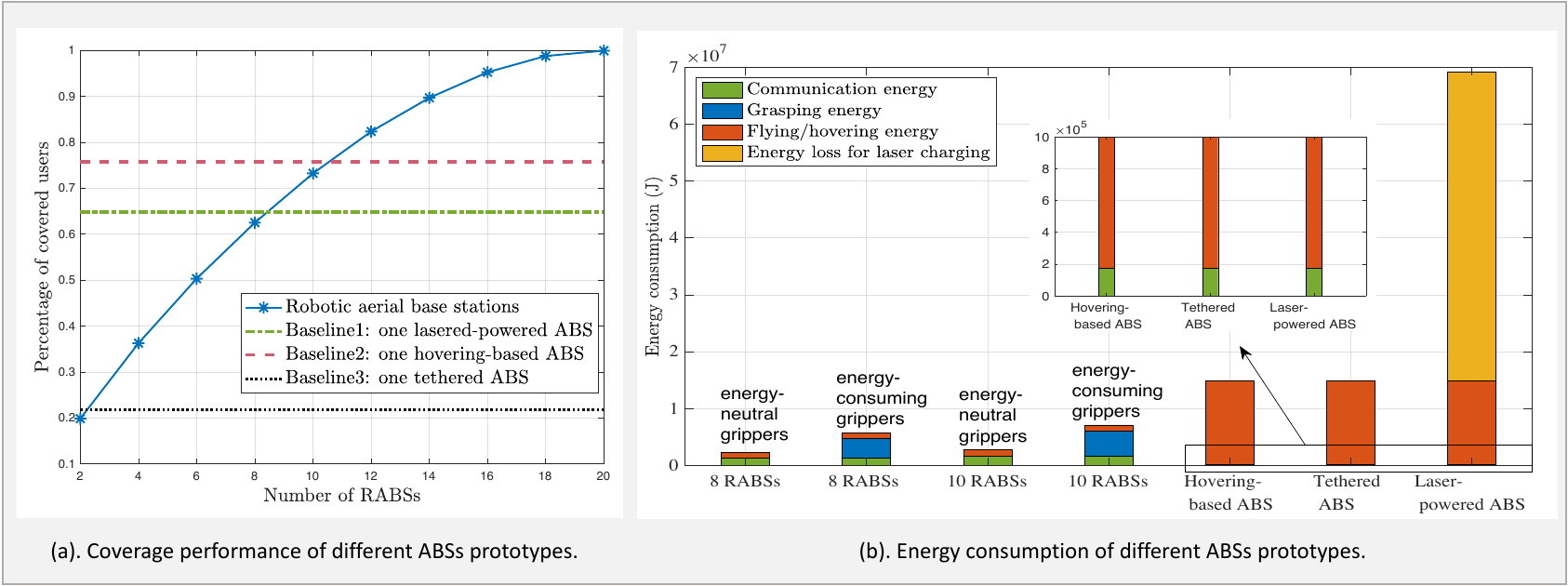}
\caption{Comparing coverage performance and energy consumption for different types of ABS.
}
\label{casestudy_1}
\end{figure*}

\textbf{\textit{1) Coverage and Energy Efficiency Improvement}}:
As reviewed in section \ref{introduction}, several novel platforms of ABS have been recently developed to overcome the endurance issue. To better clarify the benefits of the RABS pointed out above, we compare it with the following three types of ABS:  
\begin{itemize}

    \item Hovering-based ABS is the conventional ABS that can provide wireless service when hovering and change its location freely. However, one of the main challenges is that due to the limited battery capacity, hovering ABSs need to fly back to charging stations periodically and recharge via drone or battery swapping strategies \cite{lahmeri2022charging}.

    \item Tethered ABS is an ABS that is connected to a ground anchor by a physical cable or tether \cite{zhang2023deployment}. The physical link brings ABS an unlimited serving endurance, but this comes at the cost of reduced mobility, leading tethered ABSs to only serve users distributed near the ground anchor.

    \item Laser-powered ABS is another type of ABS having unlimited endurance theoretically \cite{lahmeri2022charging}. High-power laser beams transmitted from ground laser directors are used to power ABSs when hovering and flying. Although these wireless laser beams can power ABSs from a longer distance than physical tethers, the cruise range of laser-based ABSs is still limited within a ball centered around the ground laser transmitter to guarantee a safe battery level. The maximum radius of this cruise range is named as critical charging distance \cite{lahmeri2022laser}. 
    
\end{itemize}

Initially, we compare the coverage performance of these four types of ABSs. Assuming 100 users are randomly distributed over a 2 km by 2 km area, we employ the problem investigated in \cite{alzenad20173} as the simulation scenario, where ABSs are positioned to cover as many ground users as possible. Successful coverage is defined as users experiencing path loss values lower than a predetermined threshold of 118 dB. We employ the air-to-ground channel model proposed in \cite{alzenad20173} to calculate the path loss value for hovering-based, tethered, and laser-powered ABSs. However, because RABSs tend to be deployed at lower altitudes having a higher probability of being blocked, the 3GPP path loss model for urban small cells is applied for RABSs coverage \cite{3gpp2017study}. We assume that hovering-based, tethered, and laser-powered ABSs are hovering at an altitude of 100 m when providing wireless service, and the detailed constraints for different types of ABS are clarified as follows. Hovering-based ABS can be deployed freely to cover as many users as possible, which problem has been studied in \cite{alzenad20173}. Moreover, the flying range of tethered ABSs is limited by a 150 m cable linked with a ground anchor located at the origin point. Setting the laser transmission power to 800 W, the critical charging distance of laser-powered ABSs can be calculated by Eq. (10) in \cite{lahmeri2022laser}. \textcolor{black}{To simulate the grasping-based deployment for RABSs, a Manhattan-type grid map is used for the deployment of the RABS in a typical high-density urban environment, }and we assume lampposts suitable for RABS grasping are distributed every 100 m, \textcolor{black}{each with a height of 7 m (typical height for a lamppost is between 5 to 12 meters)}. RABSs can select a subset of these lampposts to grasp when providing wireless service to nearby ground users. \textcolor{black}{In future work, the study will be extended to a realistic urban layout, using city map data to model candidate anchoring distributions.}

Averaged over 100 Monte Carlo simulations, Figure \ref{casestudy_1} compares the coverage performance and corresponding energy consumption of RABSs with other baseline schemes, i.e., hovering-based, tethered, and laser-powered ABSs. 
Subfigure \ref{casestudy_1}. (a) shows that, by increasing the number of RABSs, the coverage is improved as expected. Additionally, it can be observed that a laser-powered ABS and a hovering-based ABS have almost the same performance as 8 and 10 RABSs, respectively. Subfigure \ref{casestudy_1}. (a) also shows that 20 RABSs can cover all ground users. Although compared with other ABSs located at higher altitudes, RABSs require a denser deployment to achieve a higher coverage probability, those should be considered vis-a-vis with the following result. Even a swarm of 10 RABSs still has a massive energy gain over other types of ABS.

In Subfigure \ref{casestudy_1}. (b), we compare the energy consumed by a swarm of RABSs with hovering-based, tethered, and laser-powered ABSs when operating for 24 hours. The energy consumption for RABSs includes propulsion, anchoring, and communication. We investigate RABSs that are equipped with two types of grippers, a nominal energy-consuming gripper consuming 5 W and an energy-neutral gripper. The hovering and flying power is set to 170 W and 162 W, respectively \cite{zeng2019accessing}, and the communication power is assumed to be 2 W \cite{lahmeri2022charging}. To simulate the RABSs periodical relocation, we assume that half of the RABSs need to fly 500 m every hour to update the network topology. It should be pointed out that we set the moving distance as 500 m because when a RABS needs to relocate to a further lamppost, we can always find another RABS that is closer to move to this position. Results depicted in Figure \ref{casestudy_1}. (b) show that when equipping with an energy-neutral gripper, a swarm of 8 RABSs is over 30 times more energy-efficient than a laser-powered ABS providing the same network coverage. Compared to tethered ABS and hovering ABS, a group of 10 RABSs are more than 6.6 and 5.3 times energy efficient respectively when providing the same level of wireless coverage. Besides, since the energy consumption for ABSs hovering and flying is affected by environmental factors, e.g., more energy is needed to keep stability in high winds and rainfall, in practice RABSs should show more energy gain because of their environmental independence, as illustrated in section \ref{Advantages}. Moreover, it is worth pointing out that although hovering-based ABS and RABSs cannot operate for such a long time period due to the battery capacity limitation, to ensure a fair comparison, we assume that they can be swapped and recharged to prolong the serving endurance. 
Assuming that the capacity of the onboard battery is 6700 mAh and the nominal voltage is 14.8 V \cite{lahmeri2022charging}, it can be calculated that the hovering-based ABS has to be recharged 41 times within 24 hours, while each RABS only needs to be recharged once when equipped with the energy-neutral grippers. 
\begin{figure}[!t]
\centering
\includegraphics[width=1\linewidth]{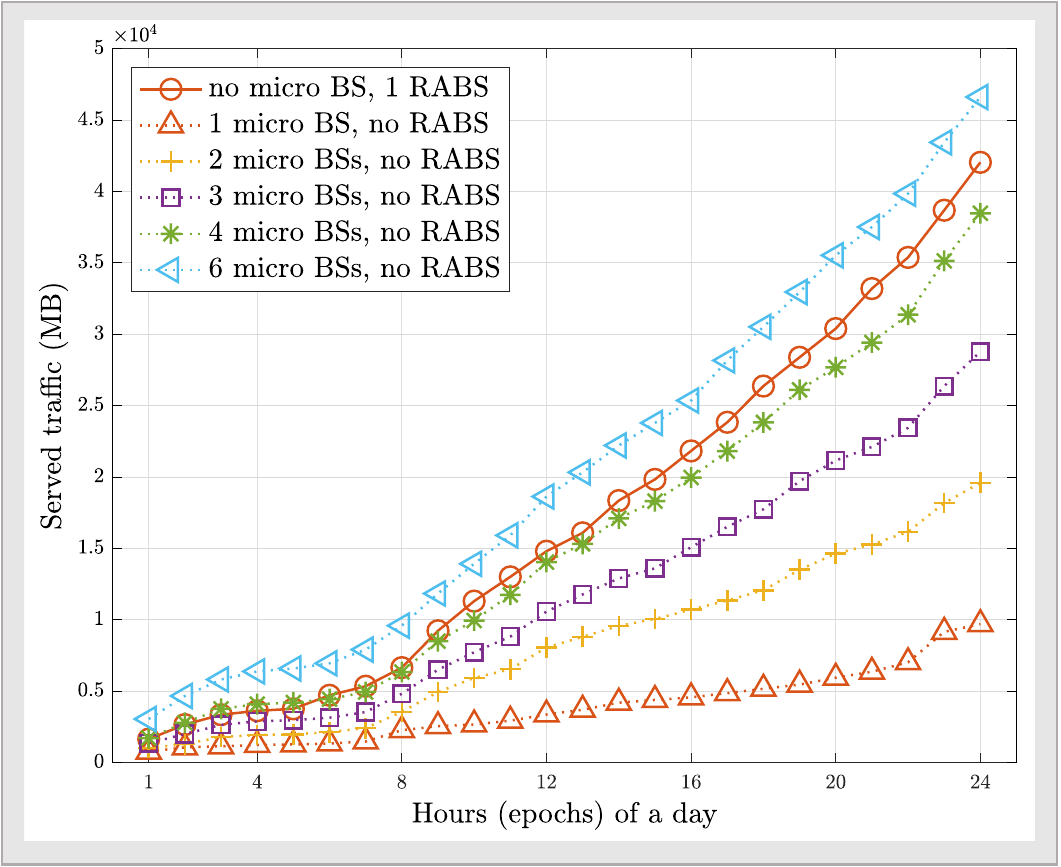}
\caption{Comparing the served traffic with fixed micro BSs.
}
\label{casestudy_2}
\end{figure}

\textbf{\textit{2) Traffic Offloading and Network Densification}}:
Next, we compare the traffic offloading performance between the RABS-assisted network and the terrestrial network with grounded micro BSs over a day-long period (i.e., 24 hours/epochs). Assuming the same scenario and propagation channel model in \cite{liao2023optimal} is investigated, and lamppost candidate locations are evenly distributed every 100 m over a 2 km by 2 km area. Traffic demand of each candidate location follows a spatial-temporal distribution model \cite{liao2023optimal}. RABS deployment is assumed to be optimized every epoch to adapt to dynamic spatial-temporal traffic demand, while micro BSs are fixed and determined greedily in terms of averaged traffic demand across the day-long period.
Average over 100 Monte Carlo simulations, Figure \ref{casestudy_2} compares the served traffic of RABS-assisted networks and terrestrial networks with micro BSs. As expected, the RABS-assisted network provides superior performance and can offload more traffic demand due to its flexibility in relocating to adapt to dynamic traffic. Numerically, one RABS can serve more than 4 times traffic compared to one micro BS, and even 1.1 times traffic of four micro BSs after 24 hours. In this regard, compared with terrestrial networks, RABS with flying capability to follow spatial-temporal traffic can efficiently deal with the unprecedented growing traffic demand of future networks, especially in scenarios with high inhomogeneity.

\section{Conclusions}
\label{Conclusion}

Robotic aerial base stations (RABSs) equipped with dexterous energy-neutral anchoring mechanisms able to grasp onto tall urban landforms can introduce significant degrees of flexibility for network densification in 6G networks. In this paper, we discuss the system architecture of RABSs in both hardware design and communication considerations by gearing two previously disconnected areas, namely non-terrestrial communications and robotic dexterous end effectors with grasping capabilities. Afterward, we articulate that RABSs could provide long-term and flexible wireless service in an energy-efficient, weather-independent, and environmentally friendly manner. Following this, we propose two case studies, one compares different types of ABSs in terms of coverage and energy performance, and the other compares ABSs to terrestrial micro BSs in terms of traffic offloading performance.


\bibliographystyle{IEEEtran}
\bibliography{IEEEabrv,reference}

\end{document}